\documentclass{INTERSPEECH2023}

\usepackage{multirow}
\usepackage{tikz}
\usepackage{subcaption}
\usepackage[tableposition=top]{caption}

\interspeechcameraready

\title{Understanding Spoken Language Development of Children with ASD Using Pre-trained Speech Embeddings}

\name{Anfeng Xu$^1$, Rajat Hebbar$^{1*}$, Rimita Lahiri$^{1*}$, Tiantian Feng$^{1*}$, Lindsay Butler$^2$, Lue Shen$^2$, Helen	Tager-Flusberg$^2$, Shrikanth Narayanan$^1$}
\address{
  $^1$University of Southern California, Los Angeles, CA, USA\\
  $^2$Boston University, Boston, MA, USA}
\email{anfengxu@usc.edu, rajatheb@usc.edu, rlahiri@usc.edu, tiantiaf@usc.edu, lbutlert@bu.edu, shenlue@bu.edu, htagerf@bu.edu, shri@ee.usc.edu}

\begin{document}

\maketitle
 \def\thefootnote{*}\footnotetext{These authors contributed equally to this work}\def\thefootnote{\arabic{footnote}}
 
\begin{abstract}
Speech processing techniques are useful for analyzing speech and language development in children with Autism Spectrum Disorder~(ASD), who are often varied and delayed in acquiring these skills.  Early identification and intervention are crucial, but traditional assessment methodologies such as caregiver reports are not adequate for the requisite behavioral phenotyping. Natural Language Sample~(NLS) analysis has gained attention as a promising complement. Researchers have developed benchmarks for spoken language capabilities in children with ASD, obtainable through the analysis of NLS. This paper proposes applications of speech processing technologies in support of automated assessment of children's spoken language development by classification between child and adult speech and between speech and nonverbal vocalization in NLS, with respective F1 macro scores of $82.6\%$ and $67.8\%$, underscoring the potential for accurate and scalable tools for ASD research and clinical use. 

\end{abstract}

\section{Introduction}
\label{section:intro}

Autism Spectrum Disorder~(ASD) is a multifaceted neurobiological developmental disorder that is characterized by impairments in social communication and interactions along with repetitive and restrictive behaviors and interests. In the USA, 1 in 44 children is reported to have ASD \cite{maenner2021prevalence}. Children with ASD typically experience significant delays in acquiring spoken language capabilities, as reported by \cite{Kim}. Timely and early identification and intervention are crucial to address the impact of ASD in children \cite{Johnson}.

Speech and language assessments have traditionally relied on standardized tests and parental reports. However, these methods are known to be inaccurate or subjective in the context of phenotyping, including language development in ASD \cite{Nordahl14}. An alternative approach that has gained attention for evaluating ASD is the use of NLS, which refers to a recording of spontaneous expressive language collected in clinically-relevant contexts and settings. NLS provides rich information about an individual's expressive language development in naturalistic settings~\cite{Barokova20}. Language samples are available within established clinical protocols for ASD diagnosis and treatment response monitoring, such as Autism Diagnostic Observation Schedule (ADOS) \cite{lord2000autism}, a semi-structured assessment administered by a trained examiner. More naturalistic language samples can be collected remotely through recordings of child-parent dyadic interactions at home \cite{Butler}. The present study uses NLS data collected from such a setting, as described in Section 3. 

ASD researchers have developed benchmarks for assessing spoken language abilities across the developmental trajectory in children with ASD, which can be determined through the analysis of NLS \cite{Tager-Flusberg}.  Continuous monitoring of children's development over time in spoken language abilities is beneficial not only for diagnostic purposes but also for assessing progress or response to treatment \cite{magiati2011autism}. However, establishing spoken language ability levels is both expensive and time-consuming, as it requires expertise and manual inspection. Robust automated audio-based measurements of spoken language capabilities sensitive to developmental changes can expedite ASD research and enhance clinical outcomes.

\begin{figure}[t]
  \centering
  \includegraphics[width=0.9\linewidth]{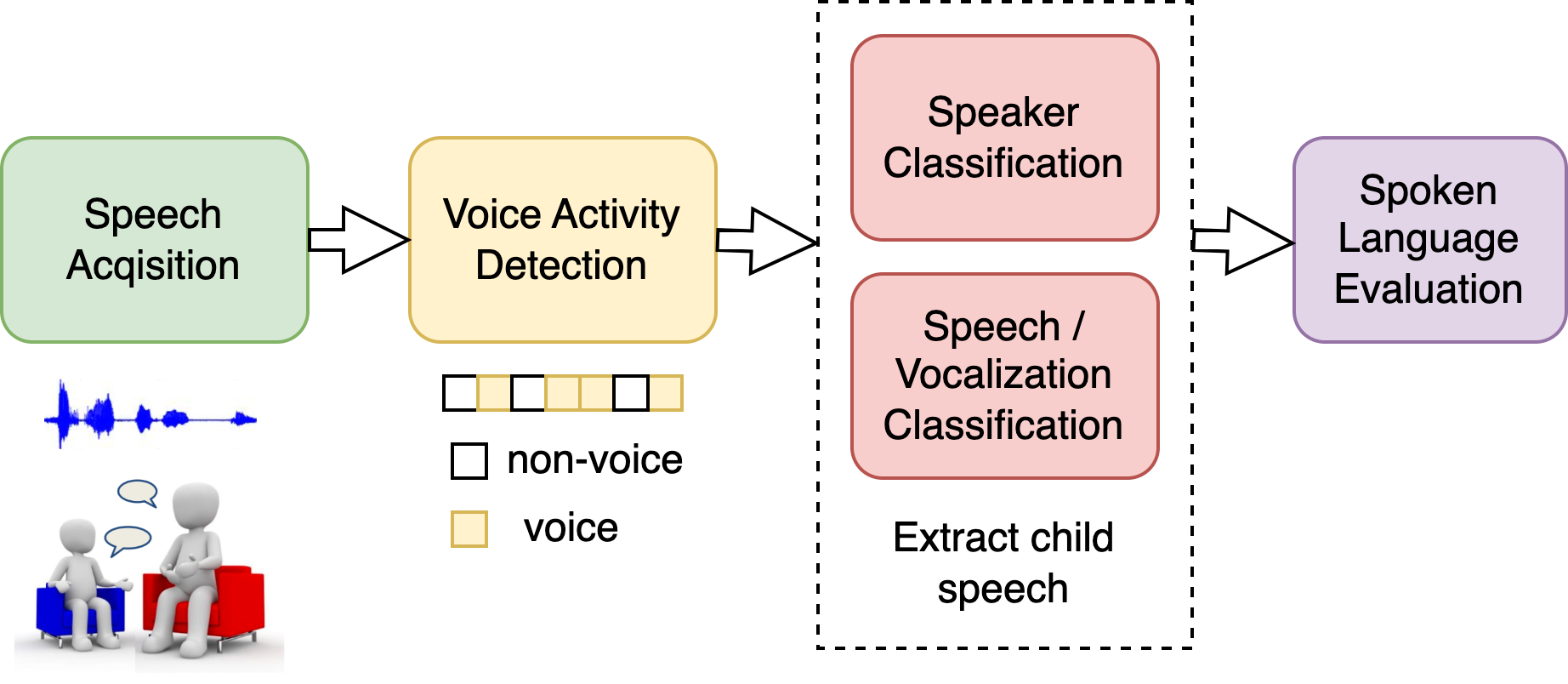}
  \caption{Spoken language assessment pipeline}
  \label{fig:speech processing}
  \vspace{-2.5mm}
\end{figure}

In this work, we propose applications of speech processing methods toward automated assessment of clinically-meaningful spoken language levels for young children with ASD. What makes the task intrinsically challenging is that the vocal signals from these young children are not well-formed, sparse, and highly variable. Indeed, the dataset we investigate in this work consists mostly of non-verbal or minimally verbal children with ASD. As a result, post speech processing techniques such as speaker diarization and automatic speech recognition (ASR) are not directly applicable or beneficial since these methods rely on the presence of adequate amounts of linguistically well-formed speech segments. On the other hand, voice activity detection (VAD) and utterance classification methods alone turn out to be critical processing steps toward automated spoken language assessment with such target subjects. 

Based on our observations from the dataset in the ASD domain, we hypothesize that the amount of speech a child produces is indicative of their spoken language levels. Thus, we focus on the child speech extraction in the spoken language assessment pipeline, as depicted in Figure~\ref{fig:speech processing}. We propose to annotate spoken language segments using four tags: \textit{intelligible speech}, \textit{unintelligible speech}, \textit{nonverbal vocalizations}, and \textit{singing}. We separate speech into \textit{intelligible speech} and \textit{unintelligible speech} because speech intelligibly is significantly impaired in minimally verbal individuals with ASD compared to verbally fluent individuals with ASD, as noted in \cite{la2020comparing} \cite{la2020comparing}. In addition, child/adult speaker segmentation is a crucial step towards the ultimate goal of automated spoken language assessment. Hence, this work focuses on classifying both speaker roles (adult/child) and spoken language (e.g. intelligible speech). In summary, the main contributions of the paper are listed below:
\begin{itemize}[leftmargin=*]
    
    \item Differing from previous literature in ASD studies that focuses on speaker classification, we introduce an additional speech and nonverbal vocalization classification task. 

    \item Based on our novel classification objective, we propose a spoken language annotation/labeling framework for automatic analysis of spoken language capabilities targeting non-verbal or minimally verbal children with ASD. 
    
    \item We show the statistical significance of spoken language labels toward our hypothesis through the ANOVA test, validating that both total utterance count and mean duration of intelligible speech utterances significantly differentiate spoken language levels (p-value $< .001$).

    \item We perform the classification utilizing three popular pre-trained speech models and report promising results in both speaker classification (Best F1: 82.6\%) and spoken language classification (Best F1: 67.8\%), demonstrating the potential of extending the current framework for future studies in ASD.

    
\end{itemize}

\section{Background}
\label{section:background}

\subsection{Speech Processing for ASD}
Automated understanding of adult-child interactions supports a variety of applications including child-centric possibilities in support of diagnosis and treatment of developmental disorders and health conditions~\cite{bone2015applying,bone2016use,sorensen2019cross}. The predominant symptoms of ASD manifest as difficulties in language and non-verbal comprehension and expression, and differences in expressive vocal prosody patterns. Thus computational analysis of interactions involving children (e.g., using speech/non-speech detection followed by speaker diarization) can serve as a valuable tool in supporting early diagnosis and intervention using vocal audio data. Prior ASD works on automated speech processing have focused on speaker diarization using spectral features and i-vectors~\cite{cristia2018talker, najafian2016speaker, zhou2016speaker}. In more recent years, ASD studies have started to investigate the use of deep neural representations to perform speaker diarization. For example, authors in \cite{xie2019multi} demonstrated that x-vectors \cite{8461375} contain more discriminative information to identify child speech utterances compared to conventional approaches. Recently, \cite{koluguri2020meta, lahiri2020learning} have addressed the challenges due to the within- and across-age and gender variability of children speech, by using a meta-learning and adversarial learning strategy, respectively. In our work, we use more recent state-of-the-art speech embedding models compared to previous works, as described in the next sub-section.

\begin{table}[t]
\caption{Summary of the pre-trained encoders used in this study.}
\vspace{-2.5mm}
    \footnotesize
    \begin{tabular*}{\linewidth}{lcccc}
        \toprule
        
        \multirow{2}{*}{\shortstack{\textbf{Pre-trained}\\\textbf{Architecture}}} & 
        \multirow{2}{*}{\textbf{Input}} & 
        \multirow{2}{*}{\shortstack{\textbf{\#Layers}}} &
        \multirow{2}{*}{\shortstack{\textbf{Hidden}\\\textbf{Size}}} & 
        \multirow{2}{*}{\shortstack{\textbf{\#Params}}}  \\ 

        & & & & \\ 
         
        \midrule
        \textbf{W2V 2.0 Base} & Raw Wave & 12 & 768 & 95.04M \\ 
        \textbf{WavLM Base+} & Raw Wave & 12 & 768 & 94.70M \\ 
        \textbf{Whisper Base} & Mel-Spec & 8 & 512 & 20.59M \\ 
        \bottomrule
    \end{tabular*}
\vspace{-0.5mm}
\label{table:pretrained_models}
\end{table}

\subsection{Speech Embeddings}

The introduction of transformers \cite{vaswani2017attention} has recently accelerated the development of Self-Supervised Learning~(SSL) techniques, the aim of which is to learn general data representations without labels. Notably, methods in speech processing such as wav2vec 2.0 \cite{baevski2020wav2vec} and HuBERT \cite{9585401} have emerged as effective approaches for learning speech representations that generalize to a variety of tasks, including phoneme classification, ASR and speech emotion recognition. We use wav2vec 2.0 \cite{baevski2020wav2vec}, WavLM \cite{chen2022wavlm}, and Whisper encoder \cite{radford2022robust} for our experiments.

Wav2vec 2.0 is a self-supervised model trained to predict masked segments of quantized speech units, which are encoded by convolutional layers. WavLM is a recently introduced self-supervised model that improves on HuBERT, which uses offline clustering to provide pseudo-labels for a masked speech prediction task. WavLM extends HuBERT by optimizing the model architecture and reformulating the reconstruction task as a masked de-noising. It achieves state-of-the-art performance on the SUPERB benchmark \cite{yang2021superb} for several downstream tasks.

Whisper is an ASR model with encoder-decoder transformer architecture trained on $680,000$ hours of diverse multilingual speech data collected from the web \cite{radford2022robust}. Since we are interested in speech representations for our application and not finetuning ASR, we extract the encoder outputs from Whisper for our experiments.

\begin{table}[t]
  \centering
  \caption{Details of the subset that was further annotated with four labels based on the audio}
  \vspace{-2.5mm}
  \footnotesize
  \begin{tabular}{lc}
    \toprule
    \multicolumn{1}{c}{\textbf{Category}} & \textbf{Statistics}  \\
    \midrule
    Age (month) & Range: 50 - 95, Mean: 79, Std: 12.3 \\
    Gender & 38 males, 7 females \\
    Count per LL & 14 (LL-1), 15 (LL-2), 16 (LL-3) \\
    Number of Activities & Range: 1 - 5, Mean: 1.9, Std: 1.1 \\
    \bottomrule
  \end{tabular}
\vspace{-4mm}
\label{table:details}
\end{table}

\section{Dataset}
\label{section:dataset}

\subsection{Dataset Details}
\label{subsection: dataset details}
In this work, we report findings using a sub-dataset from \cite{Butler, butler2022fine}, consisting of 45 videos of 15-minute interactive sessions, each involving children with autism and their parents. The data are collected remotely, in an in-home setup, where parents are instructed to choose a set of pre-defined activities that would hold their child's attention. The activities consist of 13 categories, such as games, conversations, cooking, and art. The details of the participants are reported in Table~\ref{table:details}. The video sessions are first transcribed using Systematic Analysis of Language Transcripts (SALT) \cite{miller2012systematic}. Then, domain experts use SALT to categorize the spoken language level of each child into pre-verbal communication, first words, and word combinations according to \cite{Tager-Flusberg}. For the remainder of the paper, we refer to pre-verbal communication as spoken language level 1 (LL-1), first words as spoken language level 2 (LL-2), and word combinations as spoken language level 3 (LL-3) for simplicity.


\begin{table*}[!t]
\caption{mean $\pm$ std and ANOVA test results for utterances. C = Child and A = Adult.}
\vspace{-2.5mm}
    \footnotesize
    \begin{tabular*}{\linewidth}{lccccccccc}
        \toprule
        
        \multirow{2}{*}{\textbf{Category}} &
        \multicolumn{4}{c}{\textbf{Count}} &
        \multirow{2}{*}{} &
        \multicolumn{4}{c}{\textbf{Mean Duration (s)}} \\ 
          & LL-1 & LL-2 & LL-3 & F (p-val) & & LL-1 & LL-2 & LL-3 & F (p-val) \\ 
        \cmidrule(lr){1-1} \cmidrule(lr){2-5} \cmidrule(lr){6-10}
        C - Intelligible & $2.4\pm3.8$ & $38.1\pm20.8$ & $155\pm54.0$ & $\mathbf{77}$  ($\mathbf{<.001}$) & & $0.5 \pm 0.1$ & $0.9 \pm 0.2$ & $1.2 \pm 0.5$ & $\mathbf{11}$  ($\mathbf{<.001}$) \\
        C - Unintelligible & $3.7 \pm 6.8$ & $25.2 \pm 14.3$ & $23.6 \pm 24.5$ & $\mathbf{6.5}$  ($\mathbf{.003}$) & & $0.7 \pm 0.2$ & $1.0 \pm 0.3$ & $1.0 \pm 0.3$ & $\mathbf{4.7}$  ($\mathbf{.015}$) \\
        C - Vocalization & $64.6 \pm 37.9$ & $75.2 \pm 57.4$ & $48.8 \pm 33.4$ & $1.32$  ($.279$) & & $0.9 \pm 0.3$ & $0.9 \pm 0.3$ & $0.8 \pm 0.3$ & $.26$  ($.772$) \\
        \cmidrule(lr){1-1} \cmidrule(lr){2-5} \cmidrule(lr){6-10}
        A - Intelligible & $202 \pm 66.2$ & $188 \pm 44.5$ & $182 \pm 49.0$ & $.51$ ($.604$) & & $1.3 \pm 0.5$ & $1.2 \pm 0.4$ & $1.3 \pm 0.5$ & $.11$ ($.899$) \\
        A - Unintelligible & $9.3 \pm 8.6$ & $4.1 \pm 4.8$ & $2.6 \pm 3.0$ & $\mathbf{4.9}$ ($\mathbf{.012}$) & & $0.8 \pm 0.2$ & $0.6 \pm 0.3$ & $0.9 \pm 0.4$ & $1.3$ ($.280$) \\
        A - Vocalization & $42.2 \pm 31.3$ & $22.5 \pm 13.9$ & $17.5 \pm 11.1$ & $\mathbf{5.6}$ ($\mathbf{.007}$) & & $0.6 \pm 0.2$ & $0.7 \pm 0.3$ & $0.6 \pm 0.1$ & $1.9$ ($.150$) \\
        \bottomrule
        
    \end{tabular*}
\vspace{-2.5mm}
\label{table:statistics}
\end{table*}

\subsection{Annotation Protocol}
\label{subsection: annotation}
As noted in Section~\ref{section:intro}, we have annotated child and adult (parent) speech segments using four tags: \textit{intelligible speech}, \textit{unintelligible speech}, \textit{nonverbal vocalizations}, and \textit{singing}. The annotator labels the \textit{intelligible/unintelligible} speech based on the correctness of a spoken utterance, such that a properly sounding utterance belongs to the \textit{intelligible speech} class. Meanwhile, \textit{nonverbal vocalizations} include laughter, gasping, yelling, and fillers. Additionally, utterances from third parties or with substantial overlaps are labeled separately, while background noises are disregarded. Here, the third parties may be an examiner supervising the remote session or other family members, such as siblings, the other parent, or toddlers.

\section{Method}
\label{section:methods}

\begin{figure}[t]
  \centering
  \vspace{-3mm}
  \includegraphics[width=0.7\linewidth]{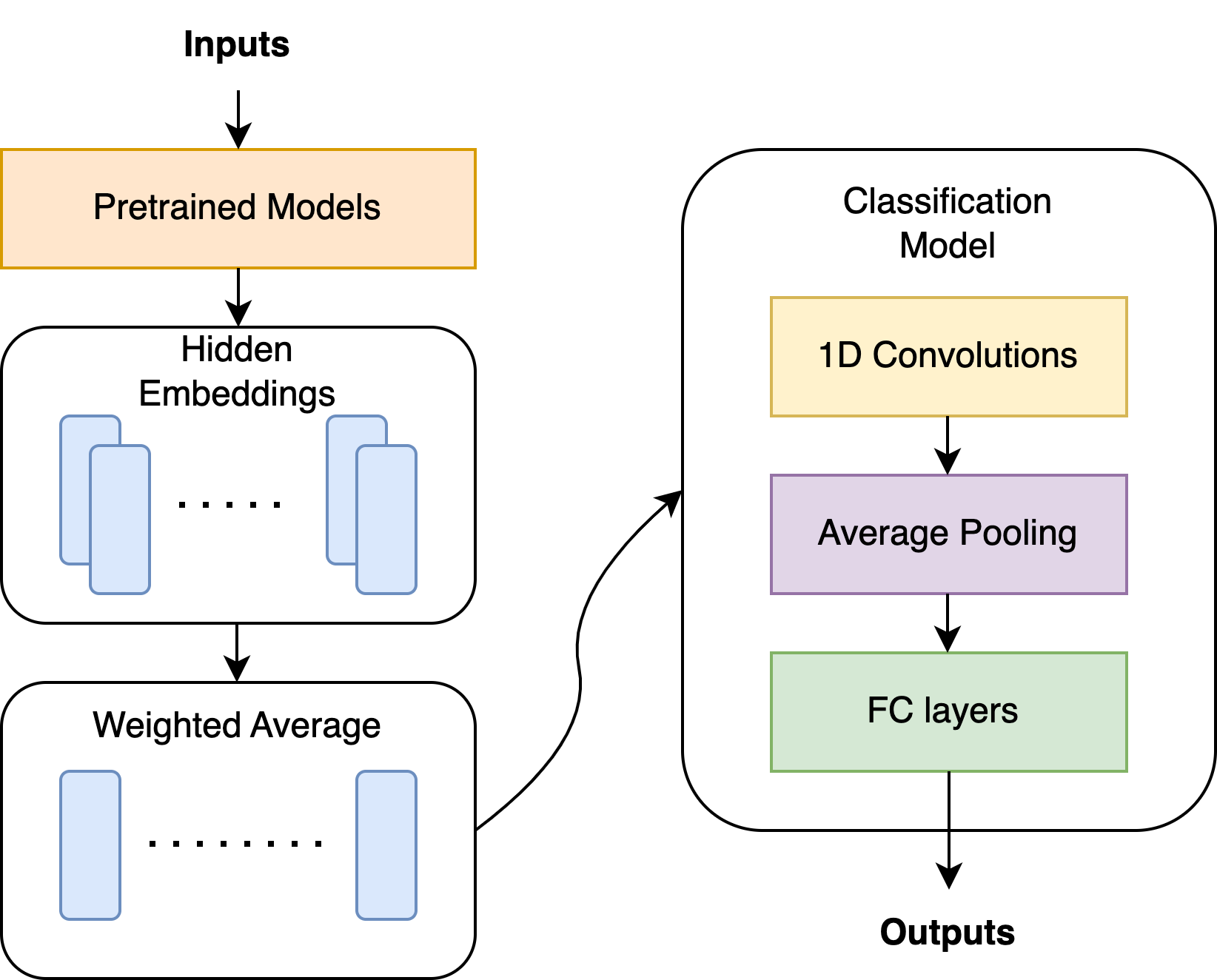}
  \caption{Audio modeling pipeline for classification}
  \label{fig:architecture}
  \vspace{-4mm}
\end{figure}

\subsection{Spoken Language Classification}
As noted in Sec~\ref{section:intro}, we hypothesize that the amount of speech from children indicates their spoken language levels. This hypothesis motivates us to design spoken language labels as described in Sec~\ref{subsection: annotation}. Statistical results to substantiate this assumption are presented in Sec~\ref{subsection:statistical_in_sl}. Thus, it is beneficial to develop a robust model to classify the attributes of both the speaker~(adult/child) and the spoken language~(speech/vocalization). Consequently, we design a modeling approach that explores speech embedding representations for the classification tasks mentioned above. Below describes the two sets of classification tasks performed in this work. 

\noindent \textbf{Child/Adult Classification}: In this speaker classification task, we include all utterances including intelligible speech, unintelligible speech, and nonverbal vocalization. 

\noindent \textbf{Speech/Nonverbal Vocalizations Classification}: For this spoken language classification task, we use the utterances from children only. We combine intelligible speech and unintelligible speech as the target speech entity for these experiments with the assumption that there is a linguistic basis for both. 

\subsection{Modeling Architecture}

We use the framework shown in Fig~\ref{fig:architecture} for both speaker and spoken language classification based on \cite{feng2023trustser}. From the pre-trained models, we extract the weighted average of all hidden layers, where the weights are learnable. Then, we feed the weighted average to a stack of three 1D convolution layers with each output channel size 256 and with Rectified Linier Unit~(ReLU) activation function. Next, mean pooling is applied over the timestamps resulting in a 256-dimensional vector. Finally, this vector is passed to fully connected layers with one hidden layer of dimension 256 and output dimension 2 for classification.

\section{Results and Analysis}
\label{section:results}

\begin{figure}[t] {
    \centering
    
    \vspace{-4mm}
    \begin{tikzpicture}
        
        \node[draw=none,fill=none] at (0,0){\includegraphics[width=0.49\linewidth]{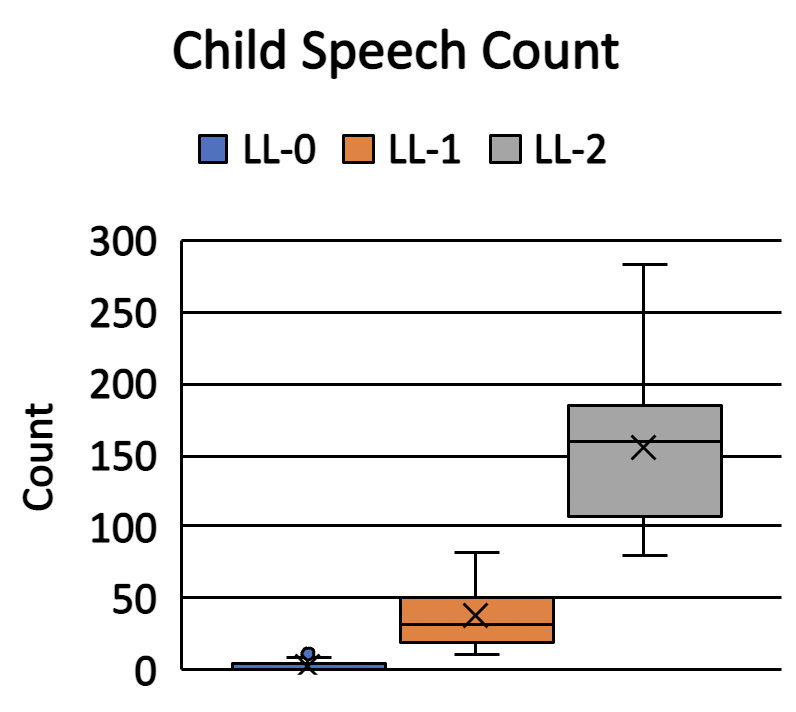}};
        
        \node[draw=none,fill=none] at (0.5\linewidth,0){\includegraphics[width=0.49\linewidth]{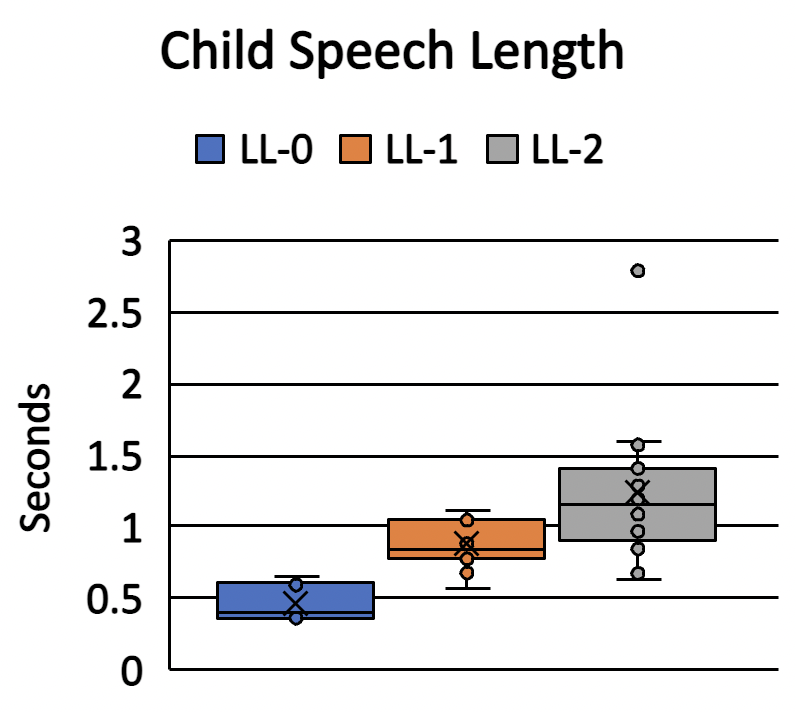}};

    \end{tikzpicture}
    \vspace{-1mm}
    \caption{Child intelligible speech utterance count and mean speech duration box plots.}
    \label{fig:boxplots}
    \vspace{-7mm}
    
} \end{figure}

\subsection{Statistical Patterns in Spoken Language}
\label{subsection:statistical_in_sl}
This section presents the statistical comparisons between different categories of spoken languages. For each session, we compute the total utterance count and mean duration of intelligible speech, unintelligible speech, and nonverbal vocalization for both children and adults. We exclude sessions with a count of zero to calculate the average duration of each type of utterance. For each of the three spoken language levels, we compute the mean and standard deviation of both the utterance count and the mean duration. Additionally, we conduct a one-way ANOVA test to compare the three groups. Statistical results and comparisons are presented in Table~\ref{table:statistics}.

As observed from Figure~\ref{fig:boxplots} and Table~\ref{table:statistics}, the total 
utterance count and mean duration of child speech utterances are the most significant indicators for differentiating the spoken language levels with a strong positive correlation. The total utterance count and mean duration of child unintelligible speech utterances are also very low for LL-1, but they do not show substantial differences between LL-2 and LL-3. No statistically significant result is observed from child nonverbal vocalizations. Counts of unintelligible utterances and vocalizations from adult show moderate evidence of a negative correlation with spoken language levels.

\subsection{Training Details}
\label{section:experiments}
Utterance samples shorter than $0.1$ seconds are discarded and all samples are truncated to have a maximum duration of $3$ seconds. Huggingface \cite{wolf2019huggingface} is used to extract the pre-trained embeddings. For whisper, we use the version pre-trained only with English. We perform 5-fold cross-validation, with the train and test split at the session level. For the training, we use an 8:2 random split between training data and validation data at the utterance level. Cross Entropy Loss is employed as the loss function with weights inversely proportional to the number of samples for each class. We use Adam optimizer with learning rate of $5\mathrm{e}{-5}$ and weight decay $1\mathrm{e}{-4}$. The batch is set to $64$ for all the experiments. A dropout of probability $0.2$ is applied for each convolutional layer. The maximum number of epochs is set to 40, with an early stopping of patience 5. The hyper-parameters are empirically determined for the best results. We use a single NVIDIA GeForce GPU 1080 Ti for the training and the whole process is expected to take less than two days. The classification pipeline is depicted in Figure~\ref{fig:architecture}.
\begin{table}[!t]
  \caption{Child / Adult Classification, F1 macro}
  \vspace{-2.5mm}
  \label{tab:child_adult}
  \centering
  \begin{tabular}{l c c c c}
    \toprule
    \multicolumn{1}{c}{\textbf{Model}}  &\textbf{LL-1}   & \textbf{LL-2} & \textbf{LL-3} & \textbf{All} \\
    \cmidrule(lr){1-1} \cmidrule(lr){2-4} \cmidrule(lr){5-5} 
    Majority Vote & $.438$ & $.365$ & $.386$ & $.395$ \\
    Wav2vec 2.0 Base & $.743$ & $.833$ & $.812$ & $.798$ \\
    WavLM Base+ & $\mathbf{.782}$ & $\mathbf{.859}$ & $\mathbf{.83}$ & $\mathbf{.826}$ \\
    Whisper Base Encoder & $.765$ & $.848$ & $.818$ & $.812$ \\
    
    \bottomrule
  \end{tabular}
  \vspace{-2.5mm}
\end{table}

\begin{table}[t]
  \caption{Speech / Vocalization Classification, F1 macro}
  \vspace{-2.5mm}
  \label{tab:speech_nonspeech}
  \centering
  \begin{tabular}{l c c c c}
    \toprule
    \multicolumn{1}{c}{\textbf{Model}}  &\textbf{LL-1}   & \textbf{LL-2} & \textbf{LL-3} & \textbf{All} \\
    \cmidrule(lr){1-1} \cmidrule(lr){2-4} \cmidrule(lr){5-5} 
    Majority Vote & $\mathbf{.696}$ & $.439$ & $.411$ & $.510$ \\
    Wav2vec 2.0 Base & $.539$ & $.694$ & $.688$ & $.644$ \\    
    WavLM Base+ & $.544$ & $.74$ & $\mathbf{.738}$ & $\mathbf{.678}$ \\
    Whisper Base Encoder & $.514$ & $\mathbf{.755}$ & $\mathbf{.738}$ & $.675$ \\
    \bottomrule
  \end{tabular}
\vspace{-2.5mm}
\end{table}

\begin{figure}[t] {
    \centering
    \vspace{-4mm}
    \begin{tikzpicture}
        
        \node[draw=none,fill=none] at (0,0.75\linewidth){\includegraphics[width=0.9\linewidth]{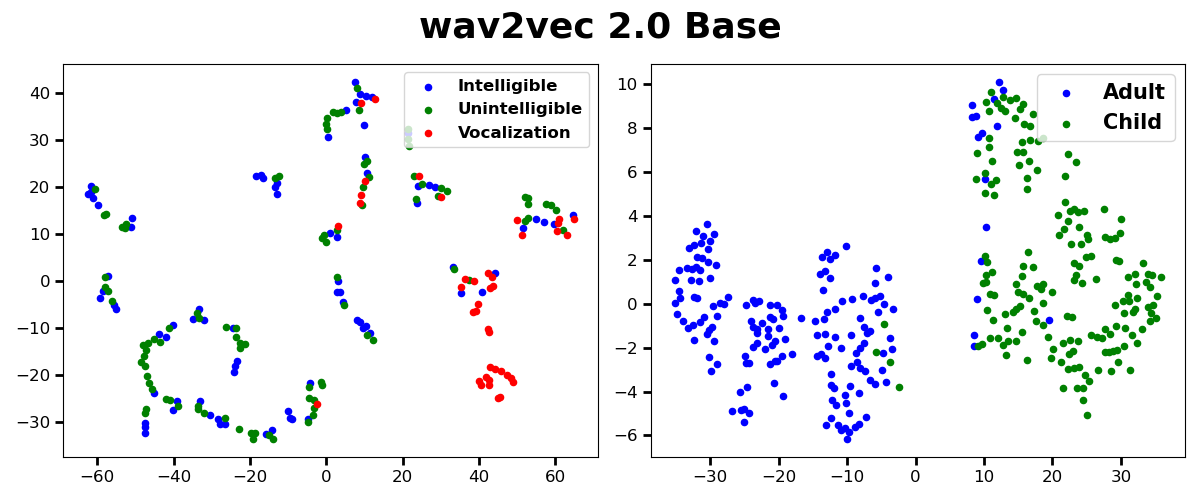}};
        
        \node[draw=none,fill=none] at (0,0.375\linewidth){\includegraphics[width=0.9\linewidth]{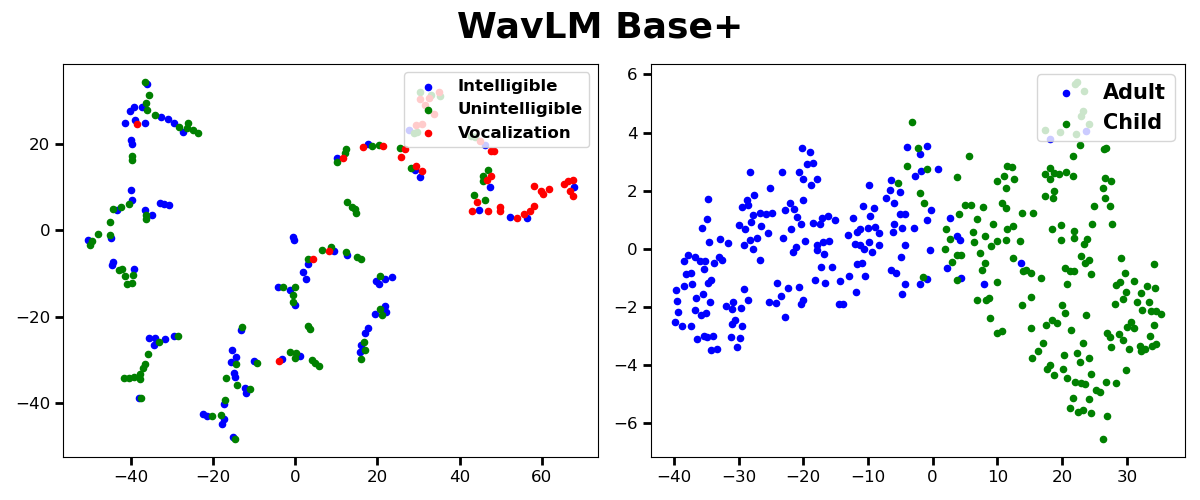}};
        
        \node[draw=none,fill=none] at (0,0){\includegraphics[width=0.9\linewidth]{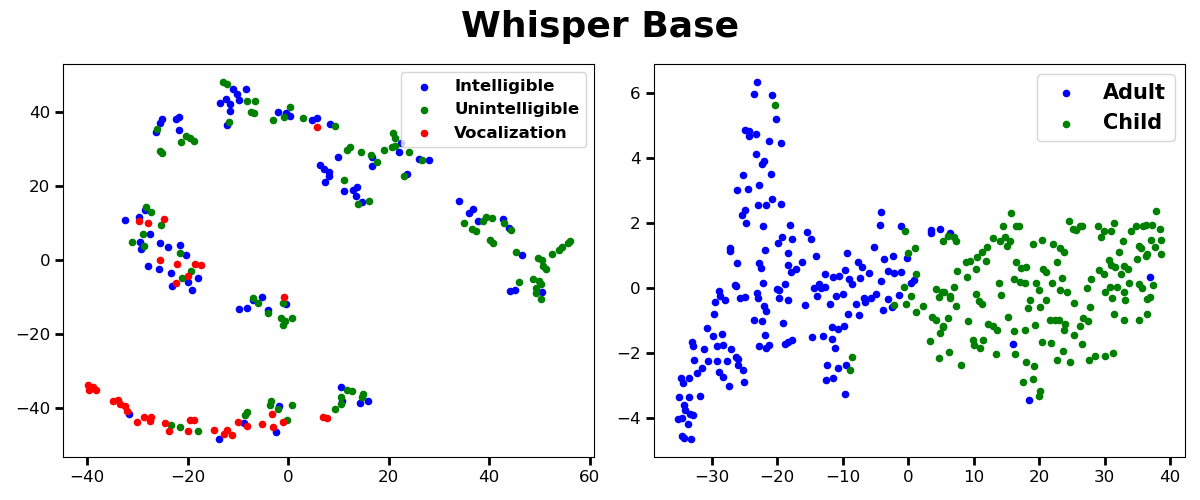}};

    \end{tikzpicture}
    \vspace{-4mm}
    \caption{T-SNE plots for each embedding.}
    \label{fig:tsne}
    \vspace{-2mm}
    
} \end{figure}
\begin{table}[t]
  
  \centering
  \caption{Female vs Male Classification F1 macro}
  \vspace{-2.5mm}
  \begin{tabular}{l c c c}
    \toprule
    \multicolumn{1}{c}{\textbf{Classification}}  & \textbf{Gender} &\textbf{LL-1}  & \textbf{LL-2, 3} \\
    \cmidrule(lr){1-1} \cmidrule(lr){2-2} \cmidrule(lr){3-4} 
    Child/Adult & Famle & $.695$ & $.794$ \\
    Child/Adult & Male & $.817$ & $.851$ \\
    Speech/Vocalization & Famle & $.554$ & $.703$ \\
    Speech/Vocalization & Male & $.541$ & $.743$ \\
    \bottomrule
  \end{tabular}
  
  \label{tab:gender}
  \vspace{-3.5mm}
\end{table}
\subsection{Modeling Results}
The results of speaker classification and spoken language classification are reported in Tab~\ref{tab:child_adult} and Tab~\ref{tab:speech_nonspeech}, respectively. We consider the majority classifier as the baseline and we report experimental findings in terms of the F1 macro score averaged over corresponding sessions.

\noindent \textbf{Speaker Classification:} Table~\ref{tab:child_adult} shows that WavLM performs the best across all three spoken language levels for speaker classification. This is likely because WavLM captures speaker information from its speech denoising modeling \cite{chen2022wavlm} during pre-training. The scores for LL-1 are lower possibly because most existing pre-trained models are trained to capture words and word combinations, while such information is sparse in LL-1. 

\noindent \textbf{Spoken Language Classification:}
WavLM and Whisper encoder perform similarly better compared to Wav2vec 2.0. Speech/vocalization classification task turns out to be a more challenging task compared to child/adult classification. The scores for LL-1 are substantially lower compared to other spoken language levels, likely due to the skewed distribution where most sessions have zero or few child speech utterances. 

\noindent \textbf{Gender-based Classification} Comparison of classification results with WavLM between males and females are shown in Table~\ref{tab:gender}. We have $4$ LL-1, $1$ LL-2, and $2$ LL-3 sessions for female children. We combine LL-2 and LL-3 since we have few female sessions for those and the results for those two levels are comparable as in Table~\ref{tab:child_adult} and Table~\ref{tab:speech_nonspeech}. The scores for child/adult classification task are notably lower for females.

\subsection{Modeling Interpretation}
T-SNE plots of three embeddings for child/adult and intelligible/unintelligible/vocalizations are shown in Figure~\ref{fig:tsne}. We use the $256$ dimensional vectors after the average pooling in the classification model. Utterances are taken from a single session with LL-3 and we use speaker and spoken language classification models trained in the cross-validation with the target session in the test split. For child/adult, we limit the utterances to intelligible speech. For intelligible/unintelligible/vocalizations, we limit the utterances only to child utterances since the model is trained only on child speech. We can clearly see the differences between child and adult as well as between intelligible/unintelligible speech and vocalization for all models.

\section{Conclusion}
\label{section:conclusion}
In this work, our focus is on assisting in the assessment of developmentally-established spoken language abilities, codified in terms of discrete stages of language levels. To achieve this, we propose a novel annotation protocol to support this goal. We present statistical analysis to demonstrate the need for speaker classification as well as spoken language classification. We explore the advantages of pre-trained speech embeddings and provide promising experimental results in both classification tasks. Our work is limited in that we did not integrate related approaches to improve classification performance and fairness. Future work includes exploring improved, fairer, more nuanced (e.g.,  of intelligible vs. unintelligible articulation), and more trustworthy methods \cite{feng2023trust}. We plan to explore multitask classification methods to improve spoken language classification. We also plan to investigate methods for constructing an end-to-end automated spoken language assessment pipeline from VAD to speaker classification and spoken language classification.

\section{Acknowledgements}
This work is supported by funds from Apple.

\bibliographystyle{IEEEtran}
\bibliography{mybib}

\end{document}